\documentclass[
twocolumn,
]{ceurart}

\usepackage{listings}

\usepackage{amsmath}
\usepackage{amssymb}
\usepackage{enumerate}
\usepackage{graphicx}
\usepackage{color,framed}
\definecolor{shadecolor}{gray}{0.9}

\renewcommand{\sf}[1]{\textsf{\textup{#1}}}              

\usepackage{listings}
\newcommand{\mycomment}[1]{}
\usepackage{cleveref}

\graphicspath{{figures/}}

\begin{document}

\conference{2025}

\title{Using ChatGPT to refine draft conceptual schemata\\in supply-driven design of multidimensional cubes}

\author[1]{Stefano Rizzi}[%
orcid=0000-0002-4617-217X,
email=stefano.rizzi@unibo.it,
]
\address[1]{DISI - University of Bologna, Viale Risorgimento, 2, Bologna, 40136 Italy}

\begin{abstract}
Refinement is a critical step in supply-driven conceptual design of multidimensional cubes because it can hardly be automated. In fact, it includes steps such as the labeling of attributes as descriptive and the removal of uninteresting attributes, thus relying on the end-users' requirements on the one hand, and on the semantics of measures, dimensions, and attributes on the other. As a consequence, it is normally carried out manually by designers in close collaboration with end-users. The goal of this work is to check whether LLMs can act as facilitators for the refinement task, so as to let it be carried out entirely ---or mostly--- by end-users. The Dimensional Fact Model is the target formalism for our study; as a representative LLM, we use ChatGPT's model GPT-4o. To achieve our goal, we formulate three research questions aimed at (i) understanding the basic competences of ChatGPT in multidimensional modeling; (ii) understanding the basic competences of ChatGPT in refinement; and (iii) investigating if the latter can be improved via prompt engineering. The results of our experiments show that, indeed, a careful prompt engineering can significantly improve the accuracy of refinement, and that the residual errors can quickly be fixed via one additional prompt. However, we conclude that, at present, some involvement of designers in refinement is still necessary to ensure the validity of the refined schemata.
\end{abstract}


\maketitle

\section{Introduction}
\label{sec:intro}

Conceptual design is a key step in the development of data warehouse (DW) systems and multidimensional databases, since it determines their information content and, ultimately, the set of queries they can answer. The goal is to create an implementation-independent representation of one or more \emph{cubes} structured according to the multidimensional model, i.e., described in terms of measures, dimensions, and attribute hierarchies. A lot of research has been done over the last couple of decades on conceptual design of cubes, mainly distinguishing between \emph{supply-driven approaches}, where the conceptual schema is determined starting from the schema of a source databases, and \emph{demand-driven approaches}, where it is created based on the end-users' requirements. 

An advantage of supply-driven design over demand-driven design is that a draft conceptual schema can be obtained from the source schema in automatic fashion, by applying an algorithm that essentially chases the functional dependencies coded in the source schema and uses them to arrange hierarchies \cite{GR09}. Although this significantly speeds up design, the draft schema must then be refined in the light of the end-users' requirements. Refinement mainly implies the following activities \cite{DBLP:journals/ase/AntonelliBR23}:
\begin{itemize}
    \item Removing attributes that are deemed not interesting for analyses.
    \item Finding \emph{descriptive attributes}, i.e., attributes that should not be used for aggregation while being useful for analyses (e.g., the name of a customer).
    \item Discretizing attributes with dense domains to make them usable for aggregation (e.g., the weight of a product).
    \item Finding \emph{optional attributes}, i.e., attributes that are undefined for some instances of the hierarchy (e.g., the \sf{State} attribute in a geographical hierarchy that also includes non-US nations).
    \item Labeling measures based on whether the SUM operator can be used or not to aggregate them (e.g., the exchange rate of dollars to euros, which cannot be summed).
\end{itemize}
Unfortunately, these activities can hardly be automated by an algorithm because they rely on the end-users' requirements on the one hand, and on the semantics of measures, dimensions, and attributes as expressed by their names on the other. Then, they must be carried out manually by designers in collaboration with end-users.

This is a typical situation in software engineering where \emph{Large Language Models} (LLMs) may come to the rescue. LLMs have proven to be a great tool for mimicking human linguistic abilities because of their capacity to learn from large corpora, which has had a disruptive effect in a number of fields \cite{DBLP:journals/corr/abs-2308-10620}, and more specifically in software engineering \cite{DBLP:journals/corr/abs-2305-12138}.
LLM-based approaches have been recently proposed for several phases of software development, e.g., requirement elicitation, code generation, and refactoring \cite{DBLP:journals/corr/abs-2305-12138,DBLP:journals/corr/abs-2303-07839,DBLP:journals/emisaij/FillC0S24}. In particular, the experiments on using LLMs for conceptual design \cite{DBLP:journals/emisaij/FillFK23,DBLP:conf/hci/LutzeW24} showed that they can help designers with this task by producing draft solutions in a timely manner ---although some human intervention is still necessary to guarantee the accuracy of the outcomes \cite{DBLP:journals/corr/abs-2303-07839}. 

The goal of this work is to check whether LLMs can act as facilitators for the refinement of conceptual schemata of multidimensional cubes, so as to relieve designers from their role or even, if possible, let refinement be carried out entirely by end-users. The Dimensional Fact Model (DFM \cite{GR09}) is the target formalism for our study; as a typical LLM, we use ChatGPT's model GPT-4o \cite{Hariri23}, which has gained popularity for its smooth user interface and natural language generating capabilities \cite{Zhou24}.
To achieve our goal, we formulate three research questions aimed at (i) understanding the basic competences of ChatGPT in multidimensional modeling and, specifically, in the DFM; (ii) understanding the basic competences of ChatGPT in the refinement of a draft DFM schema; and (iii) investigating if the latter can be improved via prompt engineering. 

This is how the rest of the paper is organized. Following the discussion of relevant literature in \Cref{sec:rw}, we provide a detailed account of our investigation in \Cref{sec:exp}, which includes the formulation of research questions, an explanation of how we designed the experiments, and the answers to the research questions. Lastly, we conclude and discuss the main validity threats to our experiments in \Cref{sec:concl}.

\section{Related work}
\label{sec:rw}
\subsection{LLMs for conceptual design}

An experiment to use LLMs for creating specifications from requirements documents in the realm of smart devices is described in \cite{DBLP:conf/hci/LutzeW24}. The authors contend that the fundamental skill of conceptual design ---namely, choosing and defending the best option to meet the requirements--- is still lacking, but they acknowledge that LLMs are very useful in later phases of the development process, like creating class diagrams from comprehensive implementation specifications and generating source code. 

Additional experiments with ChatGPT for conceptual modeling (using, for example, UML class diagrams and E/R diagrams) are discussed in \cite{DBLP:journals/emisaij/FillFK23}. 
The authors note that ChatGPT can rapidly produce an initial draft diagram from a natural language description; nevertheless, considerable modeling expertise is still needed to improve and verify the outcomes. They therefore come to the conclusion that ChatGPT can assist specialists by simplifying conceptual design, but it is obvious that it cannot take their place.

The authors of \cite{Zhou24} describe an experiment they conducted using ChatGPT and come to the conclusion that while adding LLMs to human-driven conceptual design does not dramatically affect outcomes, it does greatly reduce the time required to complete the design by requiring fewer design steps.

In \cite{DBLP:journals/corr/abs-2306-01779}, many conceptual design concepts produced by an LLM are contrasted with a baseline of crowdsourced solutions. On average, it is shown that crowdsourced ideas are more innovative, whereas LLM-generated solutions are more practical. Remarkably, it is also shown that the LLM-generated answers resemble the crowdsourced ones better when \emph{few-shot learning} is used (i.e., the prompt is supplemented with some instances of the problem).

In \cite{Chen24}, the benefits of utilizing LLMs to improve morphological analysis in conceptual design are examined. The tests demonstrate how LLMs give designers access to interdisciplinary knowledge and facilitate the methodical dissection and analysis of design problems. For optimal outcomes, LLMs and designers should work closely together and use smart prompt engineering.

With relation to use case and domain modeling, \cite{Ali24} examines how users engage with LLMs during conceptual modeling. The primary conclusions speak to the necessity of particular prompt templates to assist users on the one hand, and the usage of a recommender to offer pertinent prompts or actions on the other.

\subsection{Conceptual design of multidimensional cubes}

The main types of methods to multidimensional conceptual design are \emph{supply-driven} (or data-driven), \emph{demand-driven} (or requirement-driven), \emph{mixed}, and \emph{query-driven}. Supply-driven methods begin by designing conceptual schemata from the schemata of the data sources (such as relational schemata); end-user requirements influence design by enabling the designer to choose which data are important for making decisions and by figuring out how to structure them using the multidimensional model \cite{DBLP:conf/ssdbm/RomeroA11}. Demand-driven techniques begin with identifying end-users' business requirements, and only then do they look into how to map these requirements onto the available data sources \cite{DBLP:journals/is/0001RSAM14}.
Mixed techniques integrate requirements-driven and data-driven methods; here, both end-user requirements and data source schemata are used simultaneously \cite{DBLP:journals/infsof/TriaLT12}. The set of OLAP queries that end-users are willing to formulate is the starting point for the creation of a multidimensional schema in query-driven approaches. These queries can be specified using SQL statements \cite{DBLP:journals/dke/RomeroA10}, MDX expressions \cite{Niemi.2001}, pivot tables \cite{Bimonte.2021}, or query trees \cite{Nair.2007}.

Multidimensional modeling techniques are reviewed in \cite{DBLP:journals/jdwm/RomeroA09}, and their cost-benefit analysis is provided in \cite{DBLP:journals/is/TriaLT17}.

\section{The investigation process}
\label{sec:exp}

As stated in the Introduction, our goal in this work is to assess the
performance of ChatGPT in the refinement of a draft DFM schema obtained by supply-driven design starting from a source relational schema. We want to evaluate its basic capabilities first, and then understand to what extent they can be improved by carefully designing the prompts. 

We take as a reference an advanced form of the DFM including, besides the basic constructs of fact, measure, dimension, and attribute, the advanced constructs of descriptive attributes, optional attributes, and additivity.
In this form, a DFM schema is a graph whose root is the \emph{fact} (represented as a box with the fact name ---e.g., \sf{SALES}--- followed by a list of \emph{measures} ---e.g., \sf{Amount}), whose other nodes are \emph{attributes} ---e.g., \sf{Product}--- represented as circles and connected by arcs representing many-to-one \emph{roll-up relationships}, i.e., functional dependencies (FDs, for instance, $\sf{Product} \rightarrow \sf{Category}$). Descriptive attributes are represented without a circle; optional attributes are dashed; a non-additive measures is represented by adding its aggregation operator to its name (e.g., \sf{ExchangeRate (AVG)}).

\subsection{Research questions}
\label{sec:rqs}

We formulate the following research questions:
\begin{enumerate}[RQ.1:]
    \item Which competences does ChatGPT have in multidimensional modeling and in the DFM?
    \item Is ChatGPT capable of refining a draft DFM schema by (i) making attribute names more intuitive for end-users, (ii) showing additivity, (iii) finding descriptive attributes or discretizing them, (iv) finding optional attributes, (v) completing time hierarchies, and (vi) removing uninteresting attributes?
    \item Can the basic performance of ChatGPT in refining a draft DFM schema be improved via prompt engineering?
\end{enumerate}

\subsection{Experiment design}
\label{sec:expdesign}

Our experiment was planned in five stages: (1) establishing the base criteria and the technological environment; (2) selecting the input/output formats; (3) defining the prompt templates; (4) defining the test cases; and (5) defining the criteria for assessing the outcomes. The ensuing subsections provide an explanation of these steps.

\subsubsection{Base criteria and technology}
The criteria we follow for our experiment are listed below:
\begin{itemize}
    \item \textbf{Learning}. For learning we adopt a prompt-based learning method, which may involve giving training examples as input to the LLM and is often used as an alternative to fine-tuning \cite{DBLP:conf/models/ChenYCLMV23}. Specifically, for RQ.2 we adopt \emph{0-shot learning} (which operates with no labeled examples) since it is meant to investigate the basic capabilities of ChatGPT; for RQ.3, which implies prompt engineering, we adopt \emph{few-shot learning} and provide two training examples \cite{DBLP:conf/nips/BrownMRSKDNSSAA20}.
    To further improve learning, for RQ.3 we also employ the \emph{chain-of-thought} technique \cite{DBLP:conf/nips/Wei0SBIXCLZ22}, which includes a list of reasoning steps in the examples. RQ.1 is a simple question which implies no learning.
    
    \item \textbf{Reproducibility}. The lack of reproducibility of the tests is a significant challenge when working with LLMs because of their non-deterministic nature \cite{DBLP:journals/sosym/CamaraTBV23}. The level of ``creativity'' of ChatGPT is ruled by its temperature parameter. ChatGPT documentation suggests using a temperature value of 0 to 0.2 for more focused tasks and 0.8 to 1 for more creative tasks\footnote{\url{https://platform.openai.com/docs/api-reference/making-requests}}. In principle, no creativity is required for refining draft schemata; thus, we set the temperature to 0 for every chat.
    
    \item \textbf{Domain}. The issue domain is acknowledged to be crucial for LLMs; the more domain knowledge an LLM has, the better the model it generates \cite{DBLP:journals/sosym/CamaraTBV23}. Every example we present describes actual domains, some of which are well-known (like purchases) and others are less common (like crossfit workouts).
    
    \item \textbf{Conversation-awareness}. The answers obtained from ChatGPT may depend heavily on the previous questions asked during a conversation. Thus, as also suggested in \cite{DBLP:journals/sosym/CamaraTBV23}, we start a new chat for each case. 
    
    \item \textbf{Iteration}. In all our tests, the first answer obtained is considered. However, keeping in mind that the refinement process is inherently iterative, in RQ.3 we tried to improve the first answer by further prompting ChatGPT with suggestions.
\end{itemize}

As to the technological environment, experiments have been carried out on the ChatGPT-4o model. 
In order to facilitate a manual examination of the findings, a basic HTML+Javascript graphical user interface has also been created to illustrate and contrast the DFM schemata that ChatGPT was able to get with those found in the ground truth.

\subsubsection{Input/output format}

\begin{figure}[t]
 \centering
 \includegraphics[scale=.65]{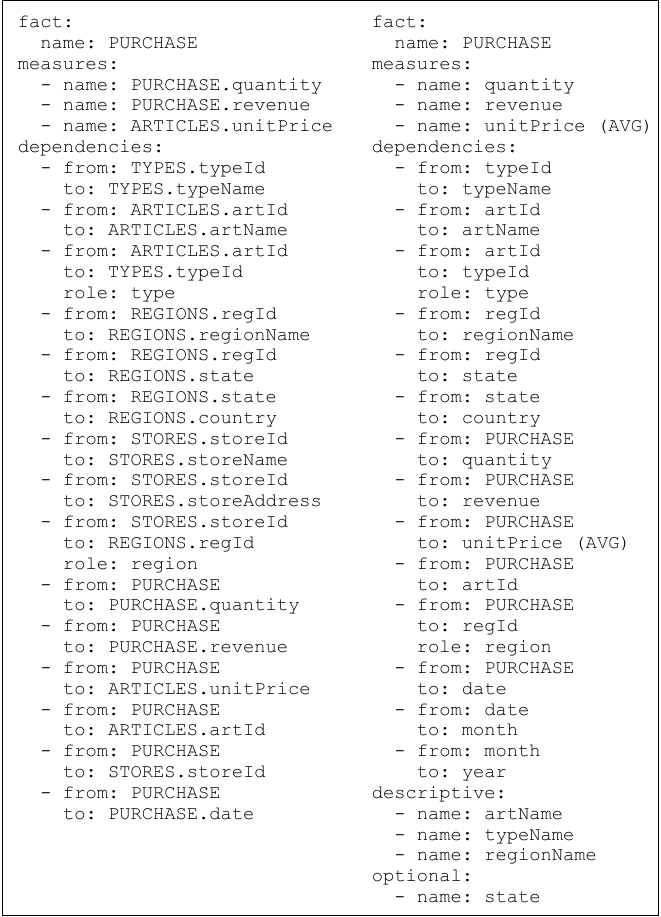}
 \caption{YAML code to represent a draft DFM schema (left) and a refined one (right)}
 \label{fig:yaml}
\end{figure}

A draft DFM schema must be provided in input for each of our research questions, and a refined one must be provided as output. We employ \emph{YAML}\footnote{\url{https://yaml.org/spec/history/2001-08-01.html}}, a human-readable data serialization language that is frequently used for configuration files and in applications where data is being saved or communicated, as a format to express DFM schemata. Since ChatGPT is familiar with YAML, it does not need any further instruction on the syntax of the language; nevertheless, it needs to be taught about the particular tags we added, specifically:
\begin{itemize}
    \item \emph{name}, to denote fact, measure, and attribute names;
    \item \emph{fact}, to denote the fact;
    \item \emph{measures}, to introduce the list of measures;
    \item \emph{dependencies}, to introduce the list of FDs;
    \item \emph{from}, \emph{to}, and \emph{role} to denote the start and end nodes of each FD and, optionally, its role in shared hierarchies;
    \item \emph{descriptive}, to introduce the list of descriptive attributes;
    \item \emph{optional}, to introduce the list of optional attributes.
\end{itemize}
Examples of the YAML code used to represent draft and refined DFM schemata are shown in \Cref{fig:yaml}.

\subsubsection{Prompt templates}
\label{sec:templates}

Following the suggestions given in \cite{DBLP:journals/corr/abs-2305-12138,DBLP:journals/corr/abs-2303-07839}, the prompts we adopt during our chats are structured according to the following templates:
    \begin{itemize}
        \item \emph{Investigation prompts}. These prompts are used in RQ.1 to assess the basic competences of ChatGPT. They are structured as simple questions, e.g., 
\begin{shaded} \footnotesize \noindent
    {\bf Prompt}: Which are the features of the Dimensional Fact Model (DFM)?
\end{shaded}
    \item \emph{Instruction prompts}. These are used in RQ.2 and RQ.3 to assign ChatGPT a task and explain how to execute it. Their structure includes:
        \begin{enumerate}
            \item ROLE: the specific roles assigned to ChatGPT and to the user to provide a context for the task; in all our tests:
\begin{shaded} \footnotesize 
    {\bf Prompt}: You are a data warehouse designer. I'm the end-user.
\end{shaded}
            This text is put in the \emph{Instructions and context} section of the chat playground.
    \item FORMAT: how the input and output (i.e., the draft and refined DFM schemata) should be coded. The format prompt we use for all tests is the following:
\begin{shaded} \footnotesize 
    {\bf Prompt}: The input I give you and the output I expect are DFM schemata in YAML formatted as follows: (1) the fact is a ``fact'' tag including a ``name'' tag; (2) all measures are listed inside a ``measures'' tag, each is an empty item containing a ``name'' tag; (3) all many-to-one associations between attributes in a hierarchy are listed inside a ``dependencies'' tag: each is an empty item containing a ``from'' tag, listing the finer attribute, a ``to'' tag, listing the coarser attribute, and optionally a ``role'' tag; (4) the ``dependencies'' list also includes an item from the fact to each dimension, and one from the fact to each measure; (5) all descriptive attributes, if any, are listed under a ``descriptive'' tag; (6) all optional attributes, if any, are listed under an ``optional'' tag. 
\end{shaded}
    \item TASK: the task assigned; for RQ.2:
\begin{shaded} \footnotesize
    {\bf Prompt (excerpt)}: You receive in input a draft DFM schema obtained by supply-driven design. A DFM schema is a connected graph where the fact is a node in which no arcs enter. [$\ldots$] 
    Your task is to refine the DFM schema given as input in different ways:
(1) Make names more intuitive for end-users.
(2) Label measures as either additive, semi-additive, or non-additive. If you find a measure is non-additive, you should add ``(AVG)'' to its name. If you find it is semi-additive, you should add ``(SUM-AVG)'' to its name. If you find it is semi-additive, just leave its name as it is.
(3) Find descriptive attributes.
(3bis) Discretize attributes.
(4) Find optional attributes. 
(5) Complete time hierarchies.
(6) Remove uninteresting attributes.
\end{shaded}
\item PROCEDURE (optional): the method suggested to perform the task; for instance, for RQ.3:
\begin{shaded} \footnotesize
    {\bf Prompt (excerpt)}: (2) Classify measures into additive, semi-additive, or non-additive. If you find a measure is non-additive, i.e., it can never be summed (e.g., the unit price of a product, a currency exchange, or a discount percentage), you should add ``(AVG)'' to its name. If you find it is semi-additive, i.e., that it can be summed along all hierarchies except temporal ones (e.g., inventory level), you should add ``(SUM-AVG)'' to its name. If you find it is additive, i.e., it can be summed along all hierarchies (e.g., the number of products sold in a day and the monthly revenue), just leave its name as it is. Make sure you rename measures also under the ``dependencies'' tag.
\end{shaded}
            \item EXAMPLE (optional): an example of some test cases, an explanation of the procedure suggested to solve it (according to the chain-of-thought principle), and the expected output. This text is put in the \emph{Examples} section of the chat playground.
        \end{enumerate}
        \item \emph{Case prompts}. These are used in RQ.2 and RQ.3 to assign a specific task to ChatGPT. Their structure includes:
        \begin{enumerate}
            \item INPUT: the input of the task (a draft DFM schema coded in YAML).
            \item TASK: the task assigned.
            \item OUTPUT: the output required (a refined DFM schema coded in YAML).
        \end{enumerate}
        For instance:
\begin{shaded} \footnotesize
    {\bf Prompt}: Here is a draft DFM schema: 
    $\langle$\emph{YAML code for draft DFM schema}$\rangle$.
    First of all, make names more intuitive for end-users. Return only the YAML without any further information/explanation.
\end{shaded}
    \end{itemize}

\subsubsection{Test cases}

We created a set of 5 test cases with increasing difficulties, based on some exercises in supply-driven design assigned to the students of a master course in Business Intelligence. Each exercise provided a source relational schema; from this schema, a draft DFM schema was created in supply-driven mode using the FD-chasing algorithm in \cite{GR09}.

\Cref{tab:cases} lists the test cases with their business domain and the number of dimensions, measures, and attributes (including dimensions) in the draft DFM schemata; the last column states whether there are shared hierarchies (i.e., nodes entered by two or more arcs, as often is the case with temporal hierarchies).

\begin{table} \footnotesize
    \centering
    \caption{Test cases used for the experiment}
    \begin{tabular}{cccccc} 
        \textbf{Case Id} & \textbf{Domain} & \rotatebox{90}{$\sharp$ \textbf{Dimensions}} & \rotatebox{90}{$\sharp$ \textbf{Measures}} & \rotatebox{90}{$\sharp$ \textbf{Attributes}} & \rotatebox{90}{\textbf{Shared Hier.}} \\\hline
        C1 & Sw Installations & 3 & 0 & 10 & N  \\
        C2 & Purchases & 3 & 3 & 12 & N \\
        C3 & Crossfit & 2 & 3 & 13 & N \\
        C4 & Car Rentals & 3 & 4 & 16 & Y \\
        C5 & Flights & 5 & 5 & 34 & Y \\\hline
    \end{tabular}
    \label{tab:cases}
\end{table}

\subsubsection{Evaluation of the results}
\label{sec:metrics}

Refinement is, to some extent, a subjective process because it largely depends on the end-user requirements. For instance, given a \sf{ProductWeight} attribute, both making it descriptive and discretizing it into \sf{WeightRanges} are reasonable refinements. As a consequence, creating a single ground truth for each test case is hardly feasible. 
So we had to proceed manually, by first identifying a set of feasible refinements for each part of each draft DFM schema, and then counting an error in the solution proposed by ChatGPT for each deviation from this set of feasible refinements.

\subsection{Answer to RQ.1: Competences}
\label{sec:rq1}

This session was aimed at assessing the understanding of the task from ChatGPT, and it only uses investigation prompts. The answers reported below are excerpted from an integration of the answers obtained at different runs.

\begin{shaded} \footnotesize
\noindent
    {\bf Prompt}: Which are the features of the Dimensional Fact Model (DFM)?

\noindent
  {\bf Answer (excerpt)}:
The DFM is a high-level conceptual model designed specifically for data warehouses and decision-support systems. Main features:
\begin{enumerate}
\item \emph{Fact Schemas} consist of fact tables and associated dimensions. Facts are quantitative data elements or measures that reflect the central focus of a business process. Each fact table is linked to dimension tables that describe the context of the facts.
\item \emph{Dimensions and Hierarchies}: dimensions provide contextual information for facts and are typically categorical attributes; each dimension can include hierarchies that define different levels of aggregation.
\item \emph{Attributes} are additional descriptive details that belong to dimensions.
\item \emph{Additivity of Measures}. Measures in the fact table can have different levels of additivity:
Fully additive allows aggregation across all dimensions;
semi-additive allows aggregation along some dimensions but not others;
Non-additive cannot be aggregated.
\item \emph{Complex Data Structures}: Degenerate dimensions, which exist as keys in the fact table without additional descriptive attributes; Snowflake schemas, i.e., extensions of dimension tables, which can be normalized to reduce redundancy, although star schemas are more common in DFM.
\end{enumerate}
\end{shaded}
The answer to this question denotes some confusion, since it mixes concepts of the DFM (attributes, measures, hierarchies, additivity) with concepts pertaining to logical design (fact tables, snowflake schemata). Besides, the definition given of non-additive measures is wrong (these measures cannot be summed but they can be aggregated using other operators).

\begin{figure*}[t]
 \centering
 \includegraphics[scale=1.2]{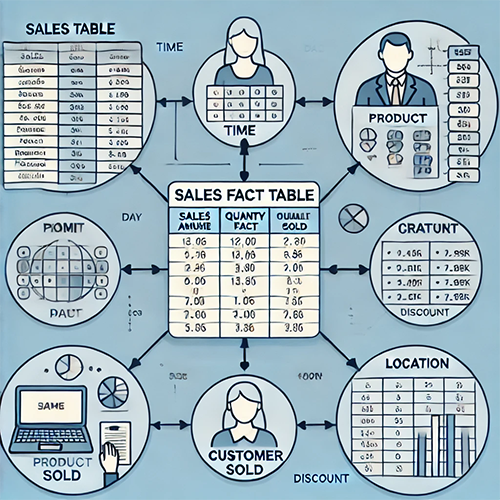} \hspace{1cm}
 \includegraphics[scale=.11]{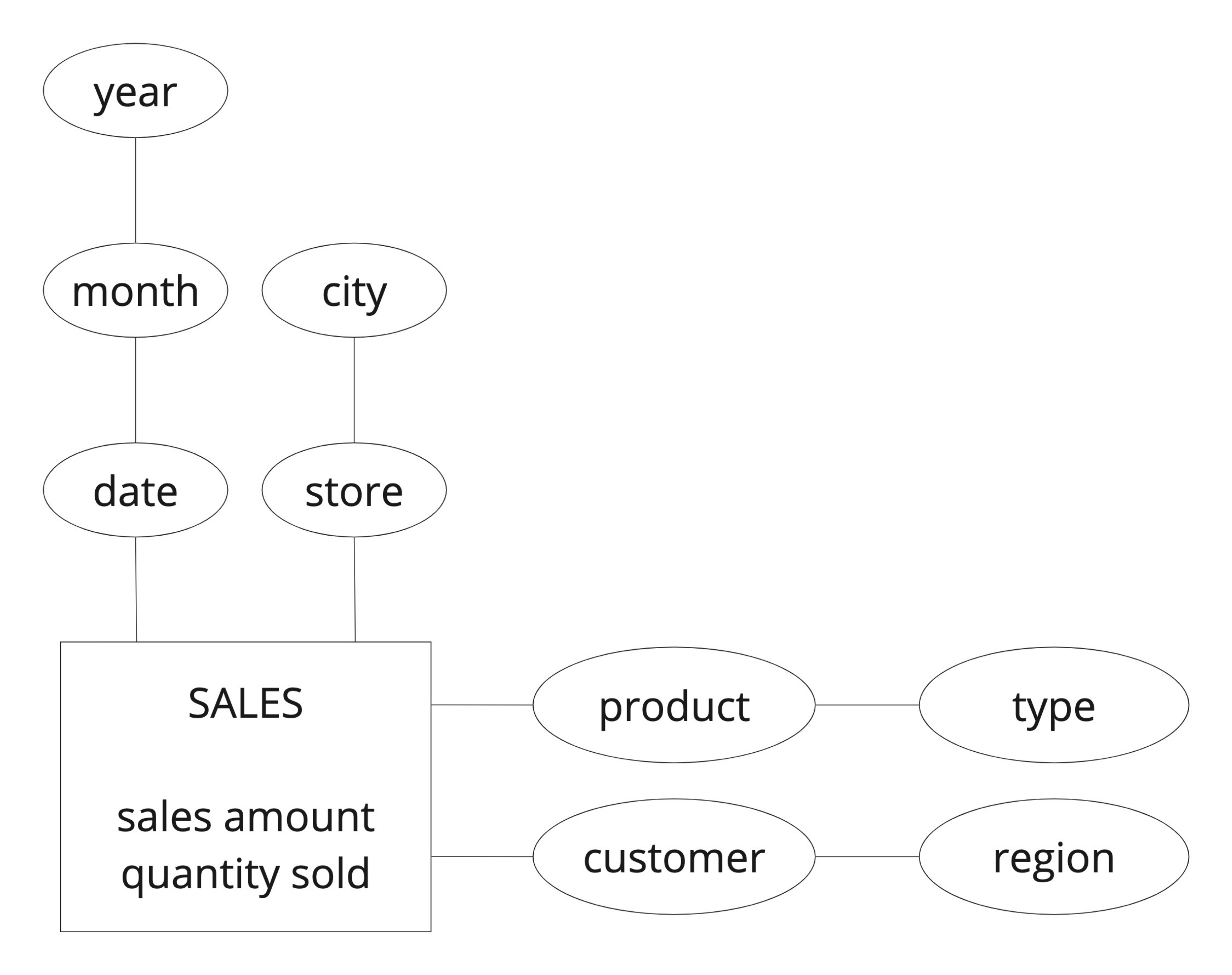}
 \caption{A DFM schema according to ChatGPT (left) and a real DFM schema (right)}
 \label{fig:fakeDFM}
\end{figure*}

\begin{shaded} \footnotesize
\noindent
    {\bf Prompt}: How would you design a multidimensional schema using the DFM?

\noindent
  {\bf Answer (excerpt)}:
Steps to draw a DFM schema:
\begin{enumerate}
\item \emph{Define the Business Process and Identify Facts}:
Clarify the objective;
Identify facts;
Determine grain.
\item \emph{Identify Dimensions and Their Attributes}:
Determine relevant dimensions;
Add attributes;
Define hierarchies.
\item \emph{Design the Fact Table}:
Create a central fact table;
Link facts with dimensions;
Define measures and specify their additivity.
\item \emph{Define Dimension Tables}:
Structure dimension tables;
Add attributes.
\item \emph{Establish Hierarchies and Aggregation Paths}:
Create hierarchies;
Define aggregation paths.
\item Draw the DFM Schema Diagram.
\end{enumerate}
\end{shaded}
An example of a \sf{SALES} DFM Schema is returned as well (\Cref{fig:fakeDFM}, left).
Again, this answer is incorrect. The terminology used mostly refers to a (logical) star schema, not to a (conceptual) DFM schema. Besides, the diagram suggested as an example (\Cref{fig:fakeDFM}, left) as no attinence to a real DFM schema (\Cref{fig:fakeDFM}, right).

\begin{shaded} \footnotesize
\noindent
    {\bf Prompt}: How would you refine a draft DFM schema obtained by supply-driven design from a source database?

\noindent
  {\bf Answer (excerpt)}: Supply-driven design often results in schemas that closely mirror the source database structure, which may not fully meet the requirements of a data warehouse optimized for decision support. Refining a draft DFM schema obtained through a supply-driven design involves aligning the schema with the business's analytical needs, optimizing it for OLAP operations, and ensuring it is intuitive for end-users.
  \begin{enumerate}
      \item \emph{Analyze Business Requirements} to ensure the schema supports the business questions it is intended to answer.
      \item \emph{Simplify and Denormalize Dimensions} to improve OLAP query performance.
      \item \emph{Redesign Fact Tables} to align them with the business process.
      \item \emph{Validate and Refine Grain} to ensure consistency.
      \item \emph{Enhance Hierarchies} by standardizing hierarchy levels and adding missing hierarchies.
      \item \emph{Enhance Semantic Richness} by replacing technical names with business-friendly terms.
      \item \emph{Optimize for OLAP} by adding summary data and creating indexes.
      \item \emph{Test with Sample Queries} to ensure it supports slicing, dicing, and aggregating as expected.
      \item \emph{Iterate and Refine} by continuously gathering feedback from business users.
  \end{enumerate}
\end{shaded}
In this answer, steps (2), (3), and (7) actually pertain to logical and physical design.

\paragraph{Discussion} Overall, it appears that ChatGPT tends to mix the conceptual, logical, and physical levels. Although it knows the DFM and its basic terminology, the descriptions it gives of the model itself and of the procedure to design and refine it, are mostly related to logical ---rather than conceptual--- design.

\subsection{Answer to RQ.2: Refinement}
\label{sec:rq2}

\begin{figure}[t]
 \centering
 \includegraphics[scale=.36]{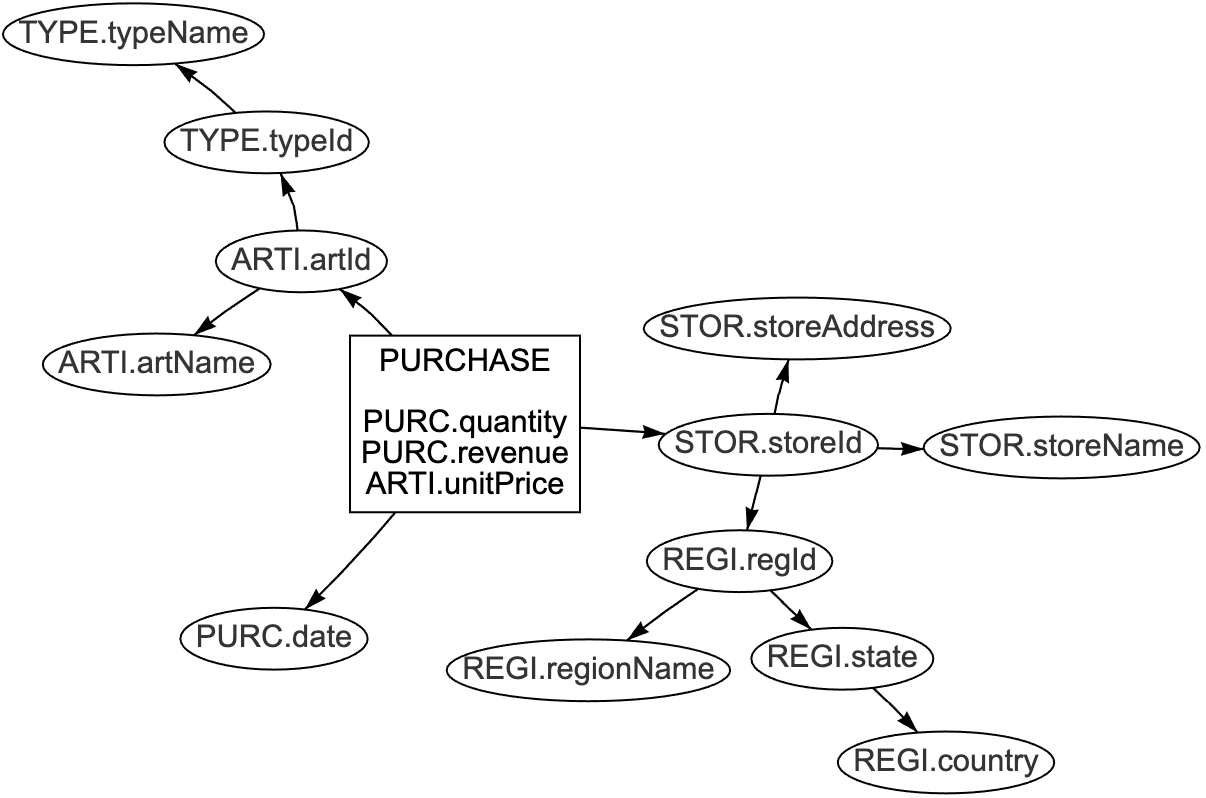}
 \caption{Draft schema for test case C2 (only the first four letters of relation names are shown)}
 \label{fig:rq2a}
\end{figure}

To put ChatGPT to the test on refinement, we fed it with the test cases of \Cref{tab:cases}. In order to enable a more precise evaluation of the abilities of ChatGPT, separate prompts are submitted for each refinement step entailed by RQ.2. Thus, for each test case we adopt a simple instruction prompt that explains the DFM constructs (see \Cref{sec:templates}) followed by a request to carry out a list of refinement steps; no PROCEDURE and EXAMPLE components are present that suggest ChatGPT how to operate. Then we formulate a sequence of case prompts that (i) specify a draft DFM schema in input, (ii) assign as a task one single refinement step, and (iii) require a refined DFM schema in output. In the following we will separately report the TASK components of each case prompt and briefly review their result.
\begin{shaded} \footnotesize
\noindent
    {\bf Prompt}: Make names more intuitive for end-users.
\end{shaded}
\noindent
The attribute and measure names in the draft DFM schema have the form \sf{RELATION\_NAME.attributeName}, being \sf{RELATION\_NAME} a table of the source relational schema used for deriving the draft schema and \sf{attributeName} one of its attributes (see \Cref{fig:rq2a}, the corresponding YAML code is the one shown in \Cref{fig:yaml}, left). 
Making these names more intuitive for end-users is mostly done correctly even if no specific procedure is suggested. In some cases the relation name was simply dropped, in others it was prefixed to the attribute name (e.g., \sf{SUPPLIER.name} became \sf{SupplierName}). In cases C3 and C4, however, the name of a compound attribute was not well chosen (e.g., \sf{WORKOUTS.date,WORKOUTS.time} became \sf{WorkoutDateTime}, while simply \sf{Workout} would have been more intuitive). In case C5, the most complex one, the shared hierarchy was mistaken and the direction of some FDs was inverted.

\begin{figure}[t]
 \centering
 \includegraphics[scale=.22]{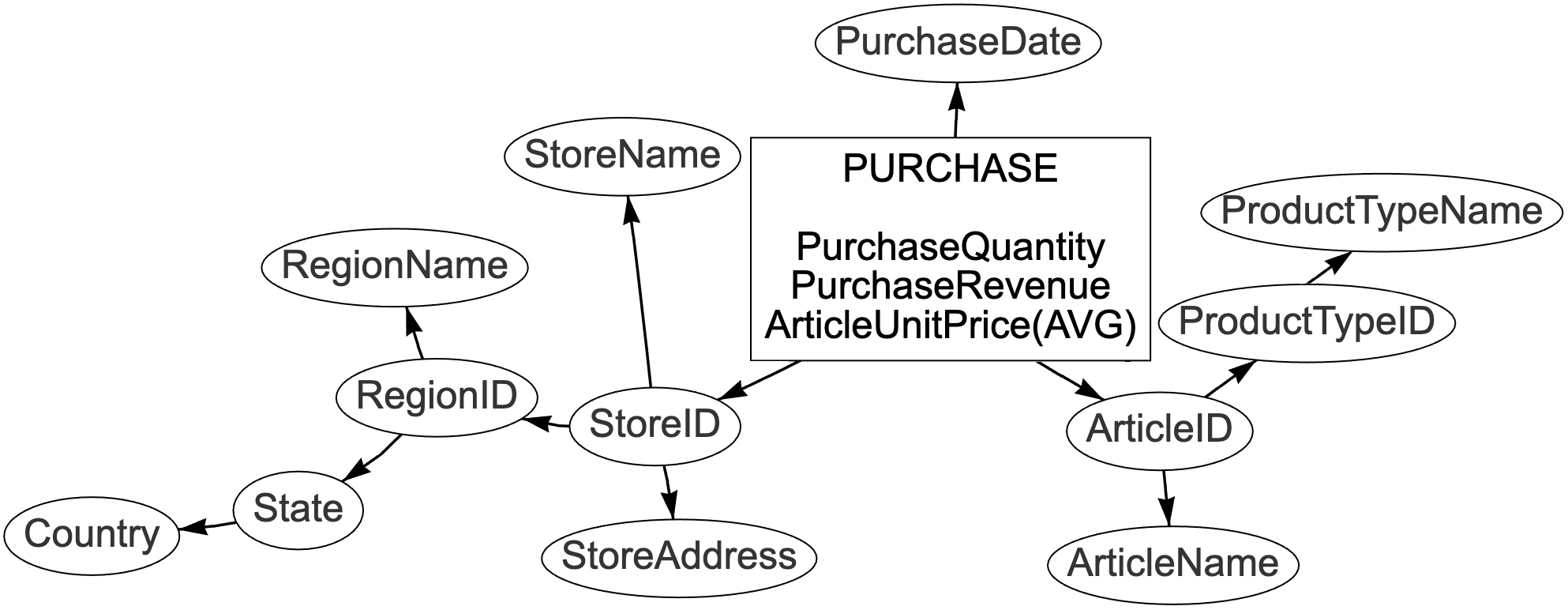}
 \caption{Partially refined schema for test case C2 with basic prompt (renaming and additivity)}
 \label{fig:rq2b}
\end{figure}

\begin{shaded} \footnotesize
\noindent
    {\bf Prompt}: Label measures as either additive, semi-additive, or non-additive.
\end{shaded}
\noindent
ChatGPT is quite good at dealing with additivity. This is surprising, considering that this task is often not easy even for end-users. The main errors we found (in C3, C4, and C5) were syntactical: the measure was renamed in the YAML code under the ``measure'' tag but not under the ``dependencies'' tag, resulting in additional fake nodes.
We only found one semantical error in C5, where \sf{NumberOfSeats} (on a flight) was labeled as semi-additive (while it should be non-additive).
\Cref{fig:rq2b} shows the DFM schema for test case C2 after renaming attributes and labeling measures.

\begin{shaded} \footnotesize
\noindent
    {\bf Prompt}: Either find descriptive attributes or discretize them.
\end{shaded}
\noindent
ChatGPT performs poorly in this task, with an average of almost four errors per test case. On the one hand, it does not know under which conditions an attribute should be made descriptive or discretized; on the other, it does not use the correct syntax as stated in the FORMAT section of the instruction prompt. Finally, in C3 and C5 it does not discretize attributes with dense domains (e.g., \sf{DepartureTime}).

The identification of optional attributes is strictly related to the end-user requirements. Thus, for this refinement step the prompt simulates an end-user statement; for instance (for C2):
\begin{shaded} \footnotesize
\noindent
    {\bf Prompt}: Not all regions have a state.
\end{shaded}
\noindent
As in the previous case, ChatGPT always fails in the syntax used (although it correctly identifies the optional attribute).

\begin{shaded} \footnotesize
\noindent
    {\bf Prompt}: Complete time hierarchies.
\end{shaded}
\noindent
Here, ChatGPT correctly adds $\sf{Month} \rightarrow \sf{Year}$ hierarchies to \sf{Date} attributes. However, it always fails in recognizing and managing shared hierarchies (in C4 and C5).

\begin{figure}[t]
 \centering
 \includegraphics[scale=.34]{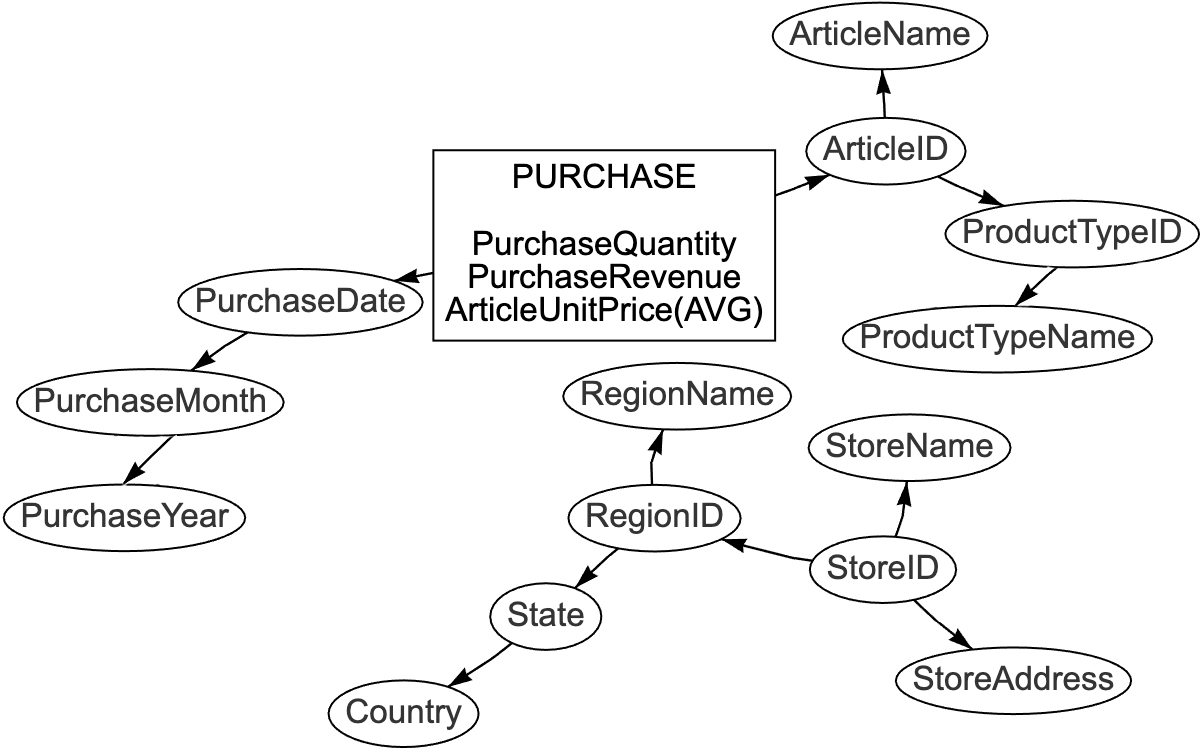}
 \caption{Refined schema for test case C2 with basic prompt (all steps)}
 \label{fig:rq2c}
\end{figure}

For the last refinement step, an indication from end-users about which attributes are deemed uninteresting for their analyses is required. Thus, like for optional attributes, the prompt simulates an end-user statement; for instance (for C2),
\begin{shaded} \footnotesize
\noindent
    {\bf Prompt}: StoreId is not interesting to me.
\end{shaded}
\noindent
ChatGPT does not know how to correctly rearrange FDs after removing an attribute, so it makes an average of 2 errors per test case for this refinement step. \Cref{fig:rq2c} shows the final DFM schema for test case C2; note that descriptive and optional attributes are not shown as such because the visualizer does not recognize the wrong YAML syntax for them, and that the graph is non-connected because some FDs were dropped.

\paragraph{Discussion} 

\begin{figure}[t]
 \centering
 \includegraphics[scale=.65]{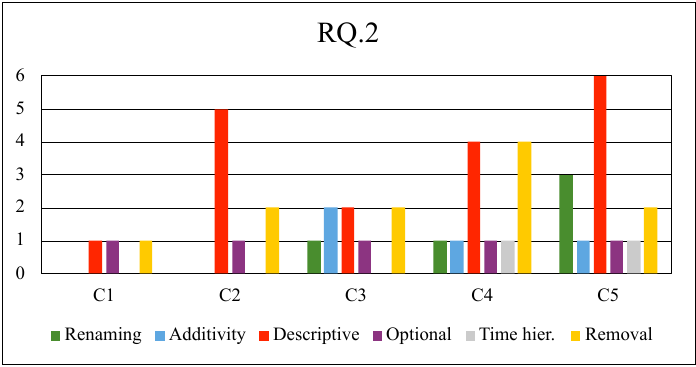}
 \caption{Number of errors in the refinement of draft DFM schemata (basic prompts)}
 \label{fig:rq2}
\end{figure}

The number of errors made at each step for each test case is shown in \Cref{fig:rq2}.
Overall, the performance are not very good but acceptable, with an average number of total refinement errors per test case equal to 9. The errors clearly tend to increase with the complexity of the draft schema; the main problems are due to the YAML syntax and to the presence of shared hierarchies. The most critical refinement steps appear to be the identification of descriptive/discretized attributes and the removal of uninteresting attributes.

\subsection{Answer to RQ.3: Improved refinement}
\label{sec:rq3}

To answer RQ.3 we incrementally crafted an instruction prompt by first trying to address the main issues emerged in RQ.2, then progressively adding specific sentences to try to fix the residual (or new) errors.
The ROLE, FORMAT, and TASK components are exactly like in RQ.2. However, for each refinement step, a PROCEDURE component is added to suggest ChatGPT how to operate. The procedure suggested to evaluate measure additivity has already been shown in \Cref{sec:templates}; as another example, here is the one suggested to remove uninteresting attributes:
\begin{shaded} \footnotesize
\noindent
    {\bf Prompt}: (6) Remove uninteresting attributes. The rule to safely remove an attribute ``b'', whose father is attribute ``a'', is to make all the children attributes of ``b'' children of ``a''. Make sure to add the arcs from ``a'' to all the children of ``b'' under the ``dependencies'' tag (even if ``a'' is the fact). If among the children of the attribute ``b'' you remove there are descriptive attributes, these should be removed as well both under the ``descriptive'' and the ``dependencies'' tags (do not remove ALL descriptive attributes, only the children of ``b''). I may ask for removing an attribute by saying that I'm not interested in it.
\end{shaded}
\begin{figure}[t]
 \centering
 \includegraphics[scale=.34]{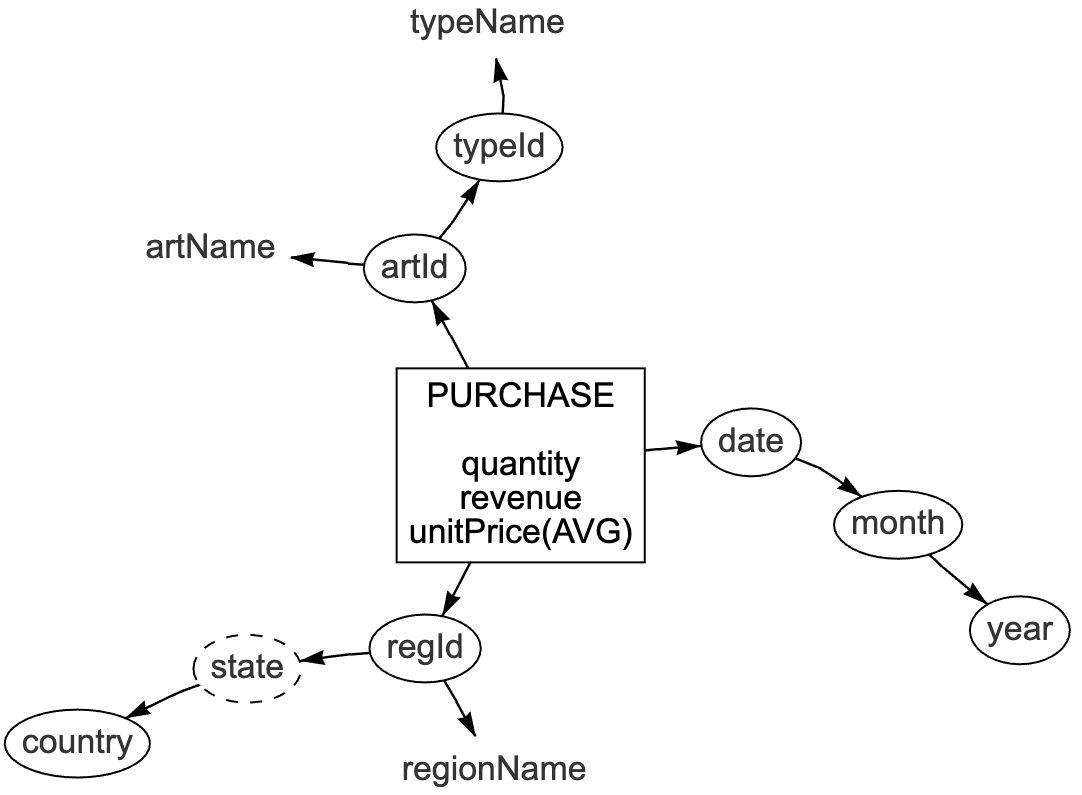}
 \caption{Refined schema for test case C2 with improved prompt (all steps); descriptive attributes are shown with no circle, optional attributes with dashed circles}
 \label{fig:rq3d}
\end{figure}
Besides, an EXAMPLE component with two examples is added to the instruction prompt. The first example has medium complexity (the draft DFM schema includes 5 dimensions, 4 measures, 11 attributes, and 1 shared hierarchy) and is structured according to the chain-of-thought principle \cite{DBLP:conf/nips/Wei0SBIXCLZ22}, which has been shown to improve learning by including a list of reasoning steps. Thus, for each single refinement step, the partial result is included together with some explanations. The second example is quite complex (7 dimensions, 1 measure, 31 attributes, and no shared hierarchies in the draft DFM schema) and only the final, refined DFM schema is included.
The case prompts are exactly the same used for RQ.2.

No errors in additivity and optional attributes are made when the improved prompt is used. A single renaming error is made in C5, due to an unrecognized shared hierarchy. A few errors are made on time hierarchies, again due to shared hierarchies. Overall, the main causes of errors are related to descriptive/discretized attributes (in some cases, a few of them are not identified) and to the removal of uninteresting attributes (sometimes, arcs are not correctly repositioned). Noticeably, all these errors could be fixed in a single iteration via specific prompts, e.g., 
\begin{shaded} \footnotesize
\noindent
    {\bf Prompt}: Merge ``drop-off date'' and ``pick-up date'' into a single ``date'' node.
\end{shaded}
\noindent
to fix a shared hierarchy.
In some cases, even generic prompts were used successfully to fix errors, e.g.,
\begin{shaded} \footnotesize
\noindent
    {\bf Prompt}: Some arcs are missing, please try again.
\end{shaded}
\noindent
to fix the FDs after removal. \Cref{fig:rq3d} shows the final DFM schema obtained for C2 after correcting two errors in descriptive attributes via an iteration prompt (the corresponding YAML code is the one shown in \Cref{fig:yaml}, right). 

\begin{figure}[t]
 \centering
 \includegraphics[scale=.65]{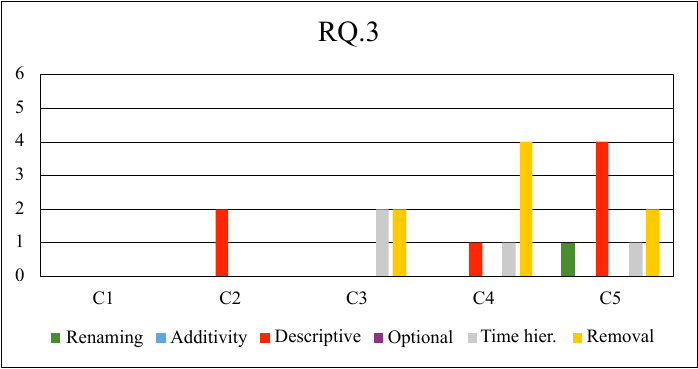}
 \caption{Number of errors in the refinement of draft DFM schemata (improved prompts)}
 \label{fig:rq3}
\end{figure}
\begin{figure}[t]
 \centering
 \includegraphics[scale=.34]{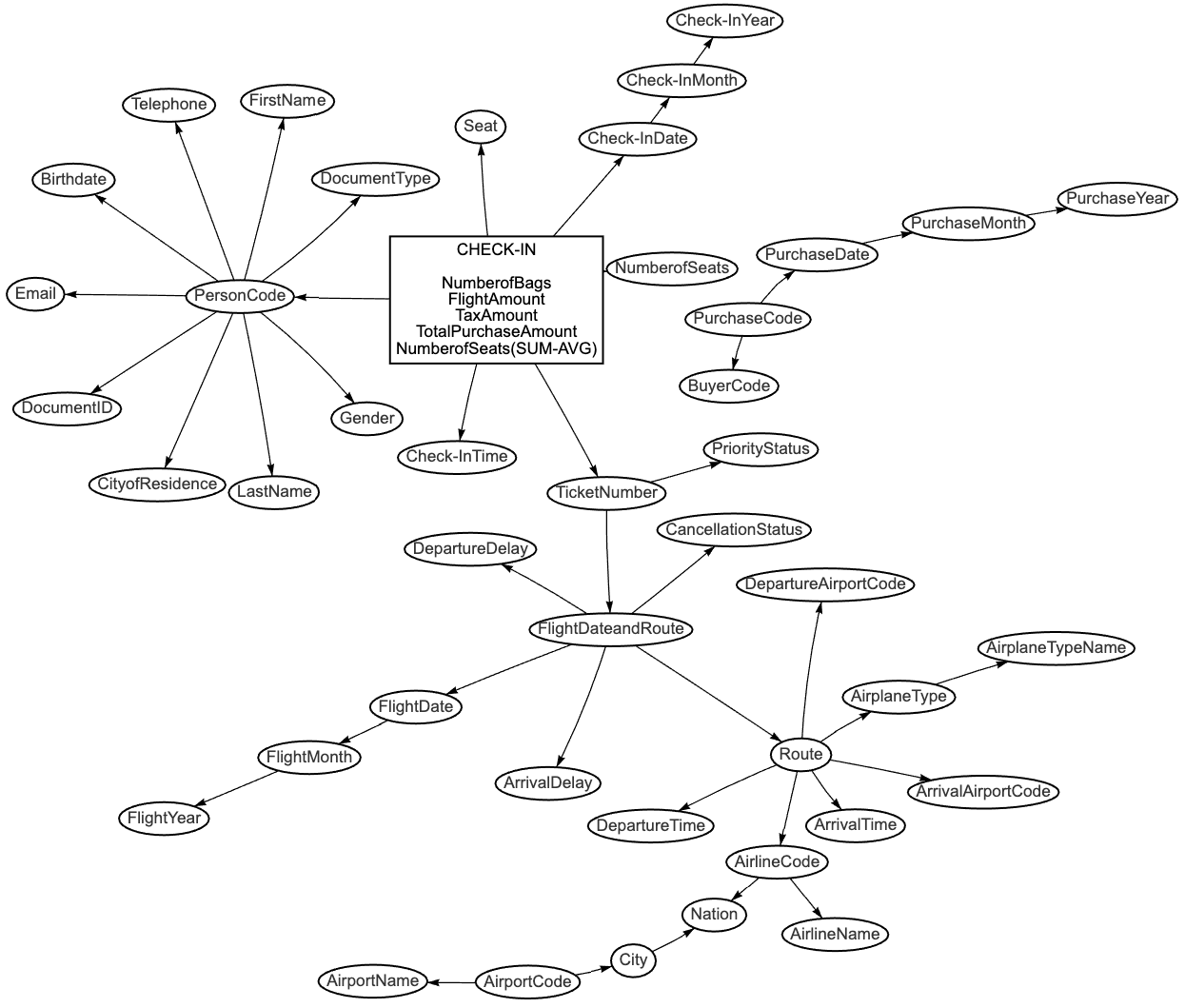}\\
 \includegraphics[scale=.34]{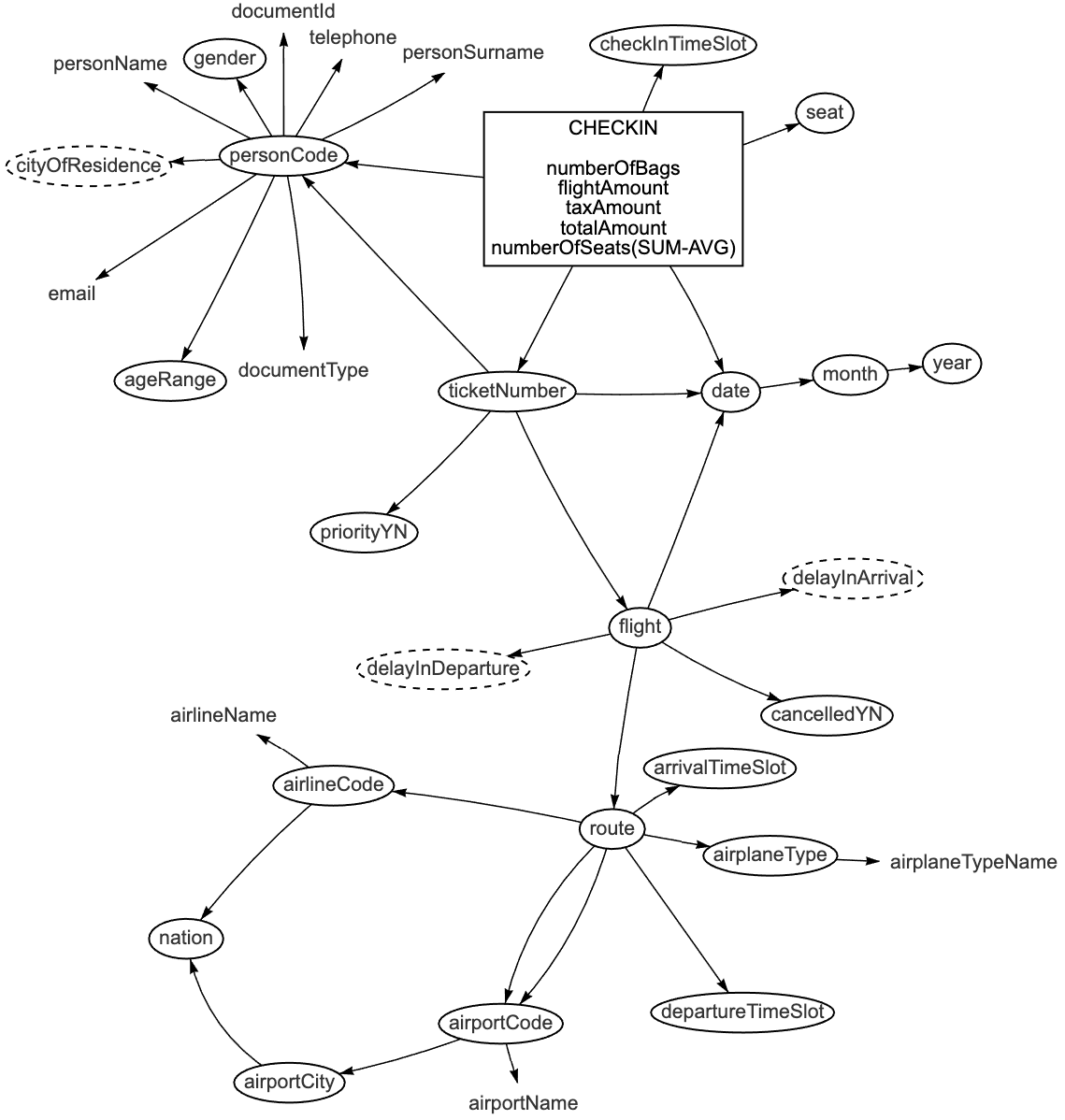}\\
 \caption{Refined schema for test case C5 with basic (top) and improved prompts (bottom)}
 \label{fig:c5}
\end{figure}
\paragraph{Discussion} 
The results, in terms of number of errors made at each step, are summarized in \Cref{fig:rq3}. It appears that prompt engineering can significantly improve the accuracy of refinement, with the average number of total refinement errors per test case decreasing from 9 to 4. The main residual errors are related to the recognition of shared hierarchies and of descriptive/discretized attributes, as well as to the removal of uninteresting attributes. In our tests, all these errors could be fixed via an additional prompt that either explains exactly how to proceed, or simply suggests to try again paying more attention to some specific aspect.
A final example is shown in \Cref{fig:c5}, which depicts the final refined DFM schemata for our most complex text case, C5, obtained using the basic and the improved prompts, respectively.

\section{Conclusion}
\label{sec:concl}

The future of software engineering is being shaped by AI, particularly through advancements in technologies like ChatGPT. These advancements are democratizing coding, making it more accessible, and boosting productivity for developers\footnote{\url{https://team-gpt.com/blog/future-of-software-engineering/}}. While knowing how to deal with LLMs can significantly help generate good code with small manual adjustments, when it comes to requirement analysis and conceptual design a considerable
modeling expertise is still needed to verify and improve the results obtained from LLMs.

\begin{figure}[t]
 \centering
 \includegraphics[scale=.7]{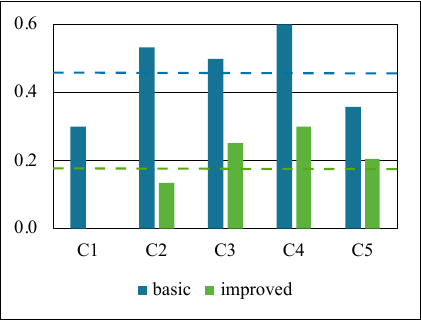}
 \caption{Average number of refinement errors per multidimensional concept}
 \label{fig:final}
\end{figure}

In this work we have investigated the capabilities of ChatGPT to cope with a specific task in conceptual design, namely, the refinement of draft DFM schemata obtained by supply-driven conceptual design of multidimensional data cubes ---a task that is normally carried out manually by designers and end-users in close collaboration. It turned out that, although ChatGPT tends to mix the conceptual level (DFM) with the logical level (star/snowflake schemata), it can provide some acceptable results on test cases with different degrees of complexity using simple prompts. Noticeably, our tests show that, when prompts are enhanced with detailed instructions and examples, the results produced significantly improve in all cases. Indeed, as shown in \Cref{fig:final}, when using an improved prompt the average number of errors per multidimensional concept across all test cases decreases from 0.5 to 0.2. In practice, the residual errors are still too many to state that no involvement of designers is necessary and that end-users can carry out refinement by directly interacting with an LLM. However, we can conclude that LLMs can significantly support designers in refinement, even considering that all residual errors in our tests could quickly be fixed via a simple additional prompt.

We close the paper by mentioning the main validity threats to our tests; addressing these threats will be the main object of our future work on this topic. The first threat is related to the reliability of the measure we use to assess the quality of refined schema. Indeed, there is a degree of subjectivity in counting the number of errors, especially because the refinement task is not deterministic and multiple ground truths could be defined for each step. Coupling this measure with objective measures, such as precision and recall for node and arc matching, could possibly address this threat. Another possible threat is related to the low statistical power of the test. To cope with this, we defined test cases with increasing sizes and different complexity; however, additional test cases that present peculiar design situations will be needed to confirm the validity of our findings.
Finally, although we set the temperature to 0 to maximize the reproducibility of the tests, some degree of variability in the results still occurs. This could be addressed by repeating each test several times and averaging the results.

\bibliography{llm}

\end{document}